\documentclass[journal]{vgtc}                     

\vgtccategory{Research}

\title{Alternatives to Contour Visualizations 
for Power Systems Data}

\author{%
  \authororcid{Isaiah Lyons-Galante}{0009-0000-4615-3721},
  \authororcid{Morteza Karimzadeh}{0000-0002-6498-1763},
  \authororcid{Samantha Molnar}{0000-0002-9589-4670},
  \authororcid{Graham Johnson}{0000-0002-4456-6909},
  \authororcid{Kenny Gruchalla}{0000-0002-8155-1175}
}

\authorfooter{
  \item Isaiah Lyons-Galante and Morteza Karimzadeh are with the University of Colorado Boulder.
  	E-mails: isaiah.lyons-galante | karimzadeh@colorado.edu
  	
  \item
  	Samantha Molnar, Graham Johnson, and Kenny Gruchalla are with the National Renewable Energy Lab.
  	E-mails: sam.molnar | graham.johnson | kenny.gruchalla@nrel.gov.
}

\abstract{%
  Electrical grids are geographical and topological structures whose voltage states are challenging to represent accurately and efficiently for visual analysis. The current common practice is to use colored contour maps, yet these can misrepresent the data. We examine the suitability of four alternative visualization methods for depicting voltage data in a geographically dense distribution system—Voronoi polygons, H3 tessellations, S2 tessellations, and a network-weighted contour map. We find that Voronoi tessellations and network-weighted contour maps more accurately represent the statistical distribution of the data than regular contour maps.
}


\teaser{
  \centering
  \includegraphics[trim={0 0 0 6cm},clip,width=\linewidth]{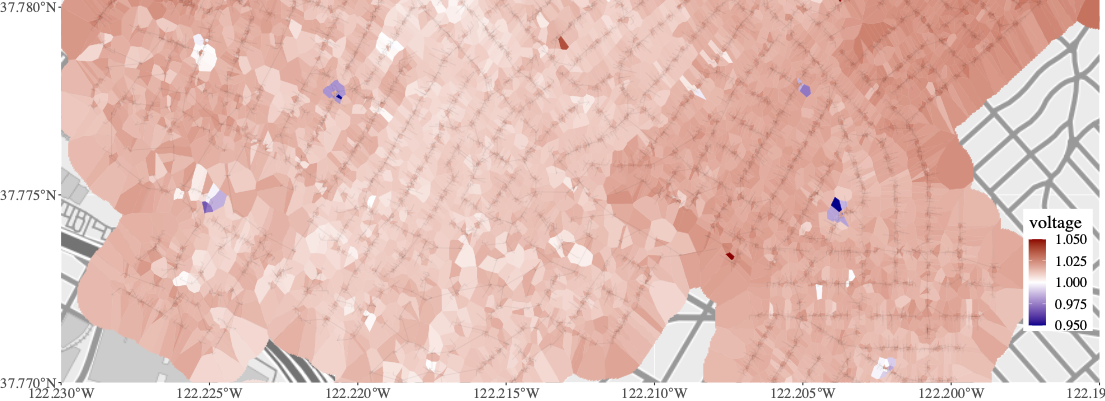}
  \caption{%
  	 Voltage on 24,000+ buses represented with Voronoi polygons that accurately depict the system voltage.
  }
  \label{fig:teaser}
}

\renewcommand{\manuscriptnotetxt}{}

\graphicspath{{figs/}{figures/}{pictures/}{images/}{./}} 

\usepackage{tabu}                      
\usepackage{booktabs}                  
\usepackage{lipsum}                    
\usepackage{mwe}                       

\usepackage{mathptmx}                  
\usepackage{amssymb}

\begin{document}


\firstsection{Introduction}

\maketitle

Ensuring a reliable power supply and maintaining voltage within acceptable limits is a regulatory obligation for electricity utility companies \cite{ANSI2020}. The voltage on each electrical junction (bus) in the grid should be within 5\% of the expected value (0.95-1.05 per unit, or p.u.) to ensure hardware functions properly. A single utility can have millions of buses to manage, necessitating effective visualization. In recent years, increased renewable energy penetration has driven power systems models to grow larger and more complex, which further drives the need to accurately visualize these systems. Integration of renewable sources and storage makes the spatial and topological relationships between network nodes are especially important because the grid is affected by both geographical and electrical phenomena such as changing weather or electric vehicle charging. Electrical grid bus voltages present a unique geovisualization problem because they are both spatial and topological, and because they have a highly variable spatial density. This makes it difficult to aggregate and visualize accurately.

\begin{figure}[t]
  \centering
  \includegraphics[trim={0 1cm 0 6cm},clip,width=\columnwidth]{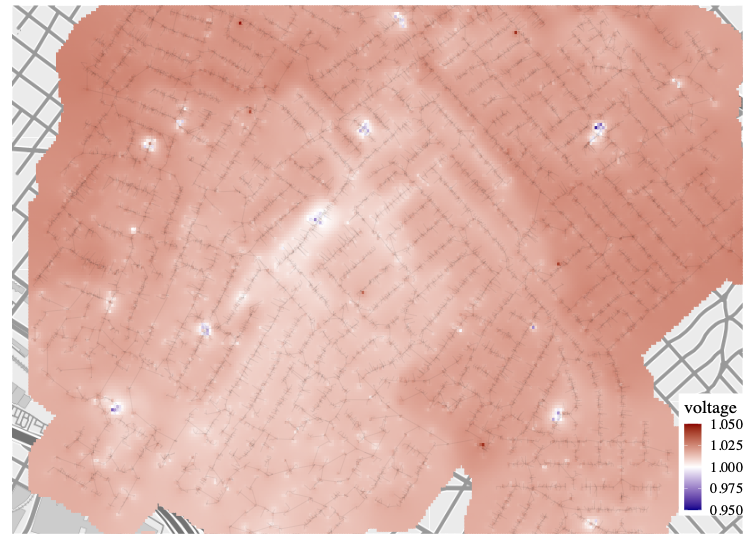}
  \includegraphics[width=\columnwidth]{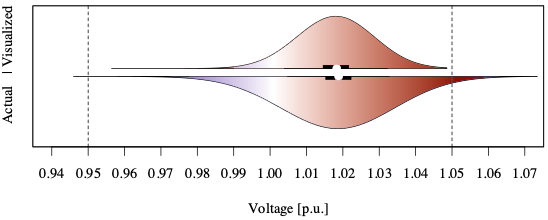}
  \caption{Voltage represented in a contour map. Values at each cell are computed with an inverse distance weighting function considering the 100 nearest neighbors. Below is a violin plot that compares the distribution of voltage values represented in the visualization (top) versus the distribution of voltage values in the actual data (bottom). The range of the distribution is reduced as values are smoothed in the contour map.}
  \label{fig:contourmap}
\end{figure}

Colored contour maps have become a standard practice for visualizations among the power systems community \cite{Gruchalla2023}. They assign color to each pixel based on a weighted-average of the voltage values of the $n$ closest buses. Each neighboring bus voltage value is weighted proportionally to the inverse distance to the pixel \cite{Wert2020}. Since every pixel is colored, this method produces a continuous distribution of bus values across a geographical area (see Figure~\ref{fig:contourmap}). They are common in research \cite{Klump2002}, operations \cite{Giri2012}, planning \cite{Kuzle2017, Overbye2021}, and commercial power system tools \cite{SiemensPSS}. However, colored contour maps have limitations in accurately representing power systems data \cite{Gruchalla2023}: the algorithm confounds the geographical and topological proximity of bus values, potentially leading to misinterpretation. Furthermore, aggregation in these maps can result in artifacts and the loss of extreme values that may be crucial for identifying critical areas. The contour map smoothing algorithm can cover up spatial discontinuities present in the data. The violin plot at the bottom of Figure~\ref{fig:contourmap} shows how the voltage data’s statistical distribution differs from that depicted on the contour map. Finally, building the contour maps can be computationally expensive for large data. This can limit the visualization systems in scale and in scope of applications.

This paper asks, are there alternative visualization methods to contour maps that can more accurately and efficiently show the state of a grid? Here, we propose alternatives to contour maps and evaluate their effectiveness in representing bus voltage in a dense distribution system. We create and compare four alternative strategies to contouring for visualizing distribution grid voltages: (a) Voronoi tessellation, (b) hexagonal tessellations, (c) four-sided tessellations with multiple resolutions, and (d) networked contours. We evaluate all methods on their effectiveness at accurately depicting the statistical distribution of voltage data,  preserving anomalous voltage values, and preserving areas of high variability. We also compare the computational efficiency of each method by using the theoretical algorithm complexity and actual compute times. Finally, we discuss challenges and opportunities with each method, and opportunities for further exploration.

\section{Related Work}

While colored contour maps have widespread use for power system data, further research has explored both contour variations and some alternatives. This includes using force-directed network layouts \cite{Wong2009}, using marker clusters in visual analytics \cite{Lawanson2018} using pseudo-geographical mosaics \cite{Overbye2019}, and varying the kernel used for contouring \cite{Wert2020}. However, each of these methods either do not represent the geography of the grid or are prone to the limitations of contouring discussed above.

\begin{figure}[t]
  \centering
  \includegraphics[trim={0 1cm 0 6cm},clip,width=\columnwidth]{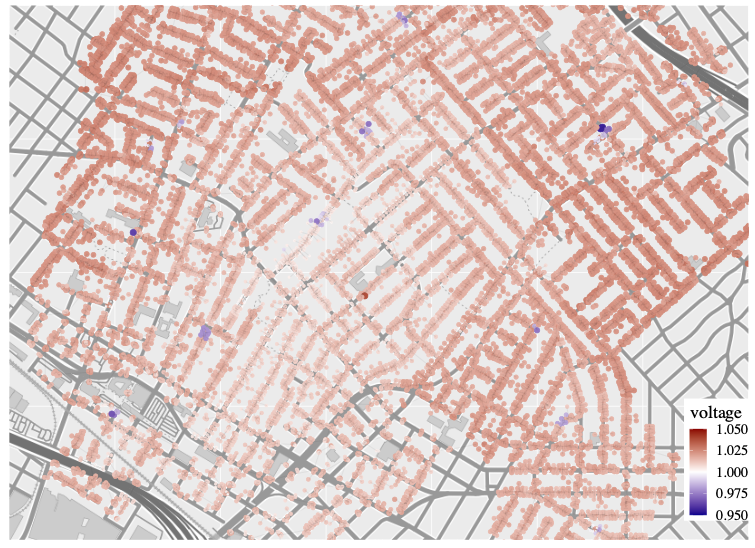}
  \includegraphics[width=\columnwidth]{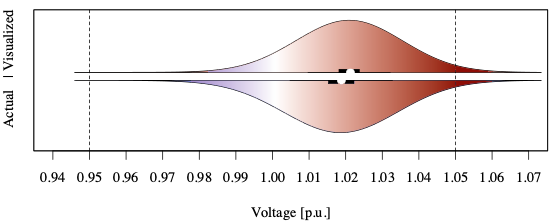}
  \caption{Voltage represented in circular glyphs. Color represents the nominal voltage value, and glyph size represents the deviation from nominal. The visualized voltage distribution is shifted away from 1.0 p.u.}
  \label{fig:glyphs}
\end{figure}

Perhaps the simplest method is directly representing the bus values with a geometric primitive (a mark or a glyph) that encodes the value by size and color ~\cite{Figueroa2020} (see Figure~\ref{fig:glyphs}). In an empirical research study, power system engineers were significantly better at identifying voltage anomalies using glyph-based representations than colored contour maps of bus values \cite{Gruchalla2023}. However, glyphs are prone to limitations on large networks. In low-density regions, the white space between glyphs can be undesirable and has the potential to be misinterpreted. In high-density regions, where the separation between buses approaches the size of a pixel, the glyphs become over-plotted and obscure some of the variation in the grid.

Tessellating is an alternative method for aggregating spatial data. A tessellation is the covering of a surface with one or more geometric shapes without gaps or overlaps. One common technique is with Voronoi polygons, where each polygon contains a single point \cite{Green1978}, in our context, a bus. Alternatively, a surface can be tessellated with a regular polygon, typically triangles, quadrilaterals, or hexagons. For geospatial tessellation, the most popular GitHub library for hexagons is H3 \cite{H3} and for quadrilaterals is S2 \cite{S2}. However, there is minimal research on the suitability of these geographic tessellations to visualize topological power systems data.

Unlike previous work building on contour maps, this paper explores alternative geographical visualizations that allow for spatial discontinuities or that consider topological relationships between nodes, while still creating a continuous covering. We also present the use statistical measures to evaluate the effectiveness of each visualization.

\section{Methods}

\textbf{Overview:} To generate our alternative visualizations, we vary both the (a) geographic tessellation shape, (b) the number of layers, and (c) the smoothing algorithm. Within (a) we try three tessellation shapes: Voronoi polygons, hexagons, and quadrilaterals. For (b), we use between 1-3 layers of tiles to show local variability while preserving a full-area covering. Within (c), we try a smoothing algorithm that considers both spatial and network distance. Finally, we evaluate the effectiveness of each with statistical measures like median, kurtosis, range, and standard deviation.

\textbf{Dataset:} The data used for the visualizations are from the SFO SmartDS synthetic model \cite{Palmintier2017}. This synthetic data was created for high-fidelity power distribution system simulations. It constitutes over 24,000 buses in a few square miles in East Oakland. The voltage data are the output of an actual model simulation. We consider the voltage on every bus at a single snapshot in time. Voltage is measured as the actual voltage divided by the expected voltage, or per unit (p.u.), with 1.0 being the expected value. We filter out outlier voltage values of 0 p.u. (about 30 buses), and we average the voltage values on buses with multiple phases (about 3000 buses). We used the same data as Gruchalla et al.~\cite{NRELData} to allow for comparisons across studies. The code used for analysis is available on GitHub \cite{GitHubRepo}. Below is a description of how each alternative visualization is created.

\textbf{Voronoi Tessellation:} A polygon is created around each bus such that all points within the polygon share that point as their nearest bus. The color of each polygon is then selected depending on the voltage value of the bus within the polygon. The vector nature of the Voronoi polygons means that the resolution of the visualization is limited only by number of pixels in the final print.

\textbf{H3 Tessellations:} We extract the H3 tile index of each bus at H3 resolutions 10-12, ranging from about $15,000m^2$ down to $300m^2$ per cell. The lowest resolution (10) has tiles large enough to ensure that the tessellated surface has no gaps while still having at least one bus in every tile. Meanwhile, the highest resolution (12) has tiles about 50X smaller, close in size to the spacing of buses in dense areas. This means that most of these tiles contain just 1 or 2 buses, which helps preserve anomalous values. The three resolutions of hexes are then colored by their mean voltage and plotted on top of one another, with the highest resolution on top. 

\textbf{S2 Tessellations:} We extract the S2 index of each bus from resolutions 16-18, ranging from about $25,000m^2$ down to about $1500m^2$. Like with H3, the lowest resolution is selected to ensure the entire space is covered with no gaps with at least one bus per tile, along with the next two higher resolutions. The highest resolution is only 16X smaller, meaning most of these tiles contain several buses. Note that S2 resolutions do not correspond 1:1 with H3 resolutions. Next, we compute the summary voltage statistics of mean and standard deviation (SD) for each S2 tile. The tiles are colored by their mean voltage, and plotted in layers, but this time with the lowest resolution on top. Then, any tile from the lower resolutions with a SD > 0.003 is filtered out. This exposes the higher resolution tiles beneath, providing more detail in the areas of high variation. Unlike with H3 tiles, the S2 tiles of higher resolutions can nest neatly inside lower resolutions. This allows for a single layer with multiple resolutions to cover the entire study area seamlessly.

\textbf{Networked Contour Map:} First, we create a network distance matrix of all nodes in the grid using Dijkstra's algorithm. We then recalculate the voltage at each bus $i$ as a weighted average of itself and its $n$ network-nearest neighbors, including itself, that are $h$ network hops away. This is expression in equation (1) below:

\begin{equation}
V_{i}^{'} = \frac{\sum_{j=1}^{n}V_{j}w_{ij}}{\sum_{j=1}^{n}w_{ij}}
\end{equation}

where $w_{ij} = \frac{1}{h_{ij}+1}$. Next, we overlay a regular grid of cells on the study area. For this dataset, the grid has about 100,000 cells, making each cell have an area of roughly $100m^2$, chosen to approximate the size of a display pixel in the final print. For each cell $x$ in the grid, the nearest $k$ bus values at euclidean distance from bus $i$ at $d_{xi}$, are averaged and mapped to a color. This is expressed in equation (2) below:

\begin{equation}
V_{x} = \frac{\sum_{i=1}^{k}V_{i}^{'}w_{xi}}{\sum_{i=1}^{k}w_{xi}}
\end{equation}

where $w_{xi}=\frac{1}{d_{xi}}$. The two key parameters here are $n$ and $k$. For $n=1$ and $k=1$, we have essentially a pixelated Voronoi tessellation. For $n=1$ and $k=100$, we have a regular colored contour map. Here, we want to emphasize network distance over spatial distance, and so we chose $n=10$ and $k=1$ for the result in the next section.

\textbf{Figures:} The buses are represented geographically as small points to illustrate their spatial relationships. Thin gray lines represent the lines that connect them to illustrate their topological relationships, all overlaid on the colored tiles. The coloring is bounded to an area within 100 meters of the grid lines to avoid showing color where there are no buses. The color scale used is blue-red diverging with white in the middle for nominal values of 1.0 per unit [p.u.]. The colors correspond to absolute voltage values to allow for identification of anomalous values and for comparability with prior research. A background street map is added for visual scale.

\textbf{Evaluation:} Visualizations methods are then compared with key statistics such as the range, median, and kurtosis. The distribution of the actual data is compared with the distribution of the visualized data using a kernel density estimate \cite{Vioplot2022} for a visual comparison of the fidelity of each visualization. Computational complexity is determined both theoretically and experimentally by recording the duration in seconds that the visualization took to create, both before and after considering the voltage data. Durations are based on a 3.1 GHz Quad-Core Intel Core i7 processor. The list of evaluation criteria are:

\begin{enumerate}
    \item Accurately depicts the statistical distribution of the real data in color space (median and kurtosis)
    \item Preserves areas of high variability (standard deviation)
    \item Highlights anomalous values (range)
    \item Is computationally efficient to create (complexity and runtime)
\end{enumerate}

\section{Results}

\begin{figure}[t!]
  \centering
  \includegraphics[trim={0 1cm 0 6cm},clip,width=\columnwidth]{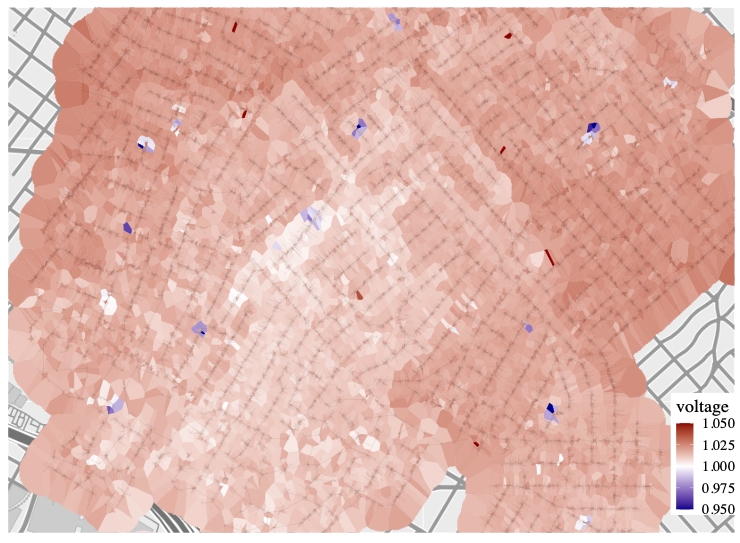}
   \includegraphics[width=\columnwidth]{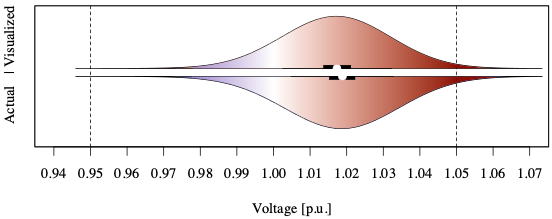}
  \caption{Voronoi tessellation showing the bus voltage value within each polygon. Each bus voltage is represented in its own polygon.}
  \label{fig:voronoi}
\end{figure}

\begin{figure}[t!]
  \centering
  \includegraphics[trim={0 1cm 0 6cm},clip,width=\columnwidth]{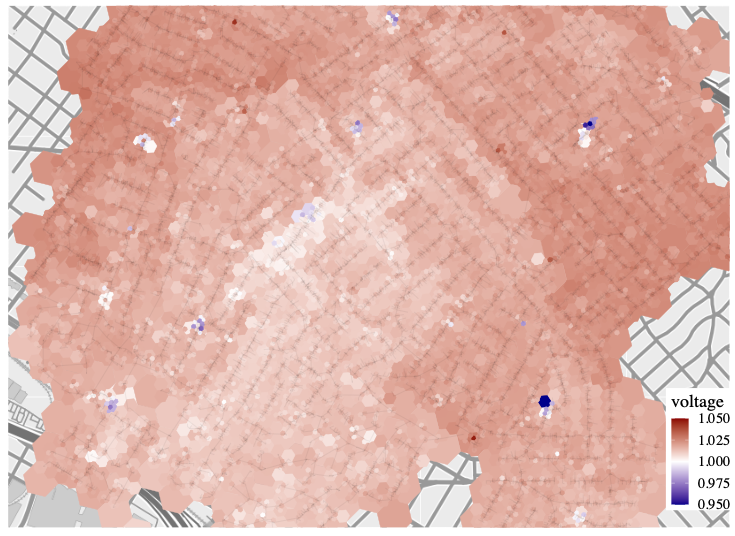}
  \includegraphics[width=\columnwidth]{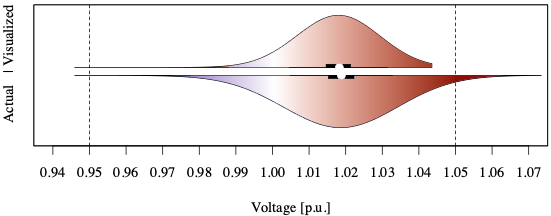}
  \caption{Hexagonal multi-layered tessellation created with H3 with each hexagon showing the mean bus voltage value of its area. High resolutions are plotted on top of low resolutions.}
  \label{fig:h3}
\end{figure}

\begin{figure}[t!]
  \centering
  \includegraphics[trim={0 1cm 0 6cm},clip,width=\columnwidth]{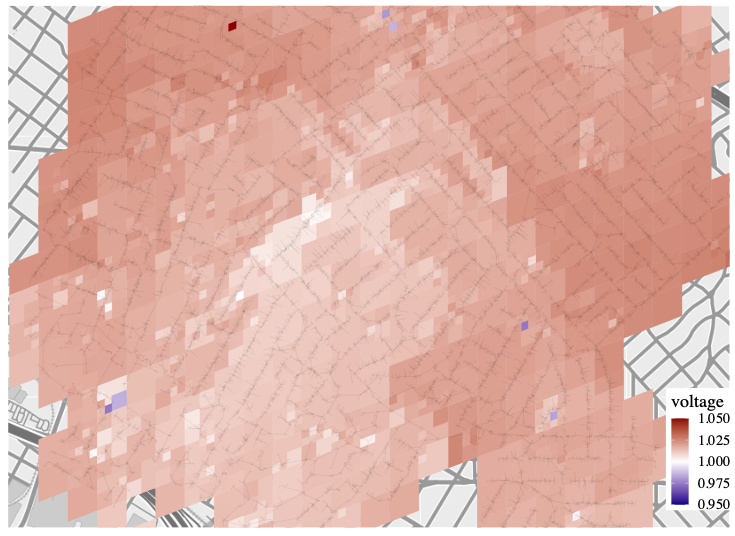}
  \includegraphics[width=\columnwidth]{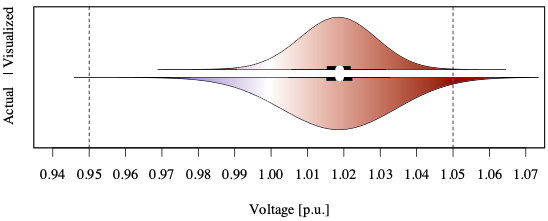}
  \caption{Quadrilateral multi-resolution tessellation created with S2 with each four-sided tile showing the mean bus voltage value of its area. Areas with high voltage variance are broken down into higher resolutions.}
  \label{fig:s2}
\end{figure}

\begin{figure}[t!]
  \centering
  \includegraphics[trim={0 1cm 0 6cm},clip,width=\columnwidth]{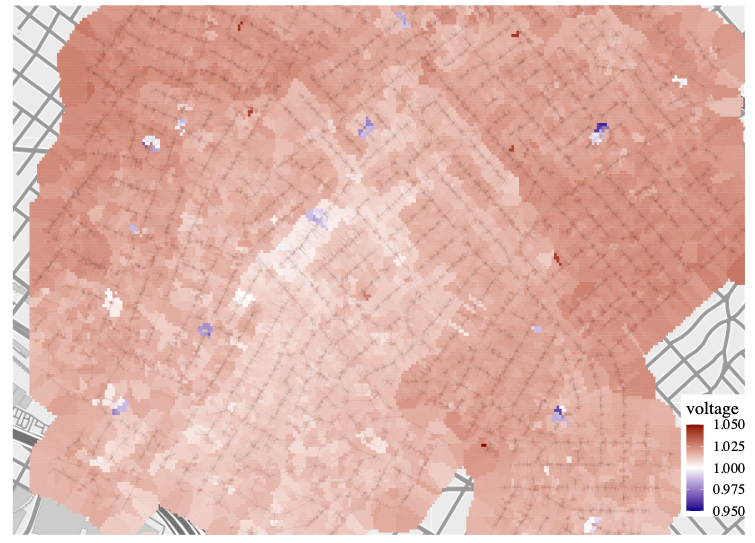}
   \includegraphics[width=\columnwidth]{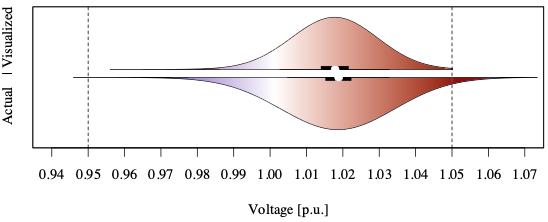}
  \caption{Networked contour map created by first taking a weighted average of the voltage value at each bus with its 10 closest topological neighbors. After voltages are recalculated, the study area with gridded with cells. Each cell is colored with the voltage value from its closest spatial neighbor.}
  \label{fig:networked}
\end{figure}

The resulting figure from each method is included above (see Figures~\ref{fig:voronoi}-\ref{fig:networked}). For each figure, the summary statistics are compiled (see Table~\ref{table:summary} for a quantitative comparison of each.  Finally, the evaluation criteria are applied to each of the visualizations with a pass or fail (see Table~\ref{table:eval}). Below, we discuss each of the resulting visualizations one at a time. We acknowledge that though we have tried to use as realistic of a dataset as possible, these results are specific to our dataset and would likely vary for a dataset with a different number of buses, different topology or geography, different voltage distributions, and for visualizations with different resolutions.

In the \textbf{Voronoi tessellation} (see Figure~\ref{fig:voronoi}), we see much more of the data variability preserved than in the contour map. Network-adjacent polygons blend together to form linear segments following the network branches, while sharp boundaries form between spatially-close but topologically-distant branches. These combine to help illustrate the topology of the grid. Outliers are not smoothed but are limited to the number of pixels in their respective polygons. The distribution mean shifts slightly left because some of the lower voltage buses are in spatially-sparse regions, leading to larger Voronoi polygons and, therefore, visual over representation.

In the \textbf{H3 tessellation} (see Figure~\ref{fig:h3}), we again see more of the data variability than in the contour map. However, some of the data’s variability is lost, as outlying buses are aggregated. The maximum actual bus value of 1.073 is averaged with an adjacent bus of 1.020, bringing the maximum hex voltage down to 1.047. Without the grid lines overlaid, the network structure is not obvious. However, the regular hexagonal tiles do create a homogeneous spatial unit for understanding the variation throughout the grid. Though the algorithmic complexity is the same as the regular contour, the computation time is significantly less because the resolution is low.  A comparably sized resolution would likely lead to the same runtime. 

In the \textbf{S2 tessellation} (see Figure~\ref{fig:s2}), we see subareas of high variation represented with more cells. However, there is influential aggregation happening, and outlier voltages are aggregated with other buses, as with H3. The four-sided cells also do not communicate the topology of the grid. The distribution tails are slightly reduced. The computation time is essentially the same as that for the H3 tessellation.

In the \textbf{networked contour map} (see Figure~\ref{fig:networked}), we can see the branches of the grid emerge in a way similar to the Voronoi contour map, and we have smoother gradients along the network lines. However, this has come at the cost of aggregating some of the outliers (while preserving others). While the statistical distribution of data is more faithful than the original contour map, the distribution tails are still reduced and anomalous buses all but eliminated. The computation time is larger because of the network distance calculations that need to be done before computing the contour map.

\begin{table*}[tbh!]
\renewcommand{\arraystretch}{1.2}
\captionsetup{justification=centering}
\caption{Summary statistics and complexity for each visualization. The "Data" column gives us the true statistics of the voltage data against which each visualization is compared. Each summary statistic corresponds to an evaluation criteria outlined in the Methods section.}
\label{table:summary}
\centering
\begin{tabular}{|l|c|c|c|c|c|c|c|}
\hline
\textbf{Metric} & \textbf{Data} & \textbf{C}ontours & \textbf{G}lyphs & \textbf{V}oronoi & \textbf{H3} & \textbf{S2} & \textbf{N}etworked Contours\\
 \hline
 \hline
 Median (p.u.) & 1.0188 & 1.0183 & 1.0213 & 1.0173 & 1.0183 & 1.0189 & 1.0179\\
 \hline
 Kurtosis & 16.632 & 4.296 & 53.810 & 13.520 & 12.655 & 10.585 & 9.203 \\
 \hline
 Standard Deviation (p.u.) & 0.0057 & 0.0048 & 0.0062 & 0.0059 & 0.0049 & 0.0049 & 0.0054 \\
 \hline
 Minimum (p.u.) & 0.9459 & 0.9564 & 0.9459 & 0.9459 & 0.9460 & 0.9689 & 0.9561\\
 \hline
 Maximum (p.u.) & 1.0734 & 1.0486 & 1.0734 & 1.0734 & 1.0468 & 1.0645 & 1.0502 \\
 \hline
 Pre-processing time (s) & NA & 350 & 0 & 150 & <10 & <10 & 430 \\
 \hline
 Rendering time (s) & NA & <10 & <10 & <10 & 40 & 30 & 30\\
  \hline
 Pre-processing complexity & NA & $O(n\, m)$ & $O(1)$ & $O(n\log{}n)$ & $O(nm)$ & $O(nm)$ & $O(n\, m)+O(n\log{}n)$ \\
 \hline
 Rendering complexity & NA & $O(m)$ & $O(n)$ & $O(n)$ & $O(m)$ & $O(m)$ & $O(m)$\\
 \hline
\end{tabular}
\end{table*}

\begin{table}[tbh!]
\renewcommand{\arraystretch}{1.2}
\captionsetup{justification=centering}
\caption{Evaluation criteria results for each visualization. A visualization receives a green "\checkmark" if the corresponding statistics are close in value to the true data, or if the rendering is relatively efficient. Otherwise, it is given a red "$x$".}
\label{table:eval}
\centering
\begin{tabular}{|l|c|c|c|c|c|c|}
\hline
\textbf{Metric} & \textbf{C} & \textbf{G} & \textbf{V} & \textbf{H3} & \textbf{S2} & \textbf{N}\\
 \hline
 \hline
 Accurate distribution & \checkmark\cellcolor{green!50} & $x$\cellcolor{red!50} & \checkmark\cellcolor{green!50} &  \checkmark\cellcolor{green!50} & \checkmark\cellcolor{green!50} & \checkmark\cellcolor{green!50} \\
  \hline
Preserves areas of high variance & $x$\cellcolor{red!50} & \checkmark\cellcolor{green!50} &  \checkmark\cellcolor{green!50} & $x$\cellcolor{red!50} & $x$\cellcolor{red!50} & \checkmark\cellcolor{green!50} \\
  \hline
Preserves tails / outliers & $x$\cellcolor{red!50} & \checkmark\cellcolor{green!50} & \checkmark\cellcolor{green!50} & $x$\cellcolor{red!50} & $x$\cellcolor{red!50} & $x$\cellcolor{red!50} \\
  \hline
Computationally efficient & $x$\cellcolor{red!50} & \checkmark\cellcolor{green!50} & \checkmark\cellcolor{green!50} & $x$\cellcolor{red!50} & $x$\cellcolor{red!50} & $x$\cellcolor{red!50} \\
 \hline
\end{tabular}
\end{table}
\section{Discussion}
Given the four criteria for evaluating the visualizations, the top performing visualization is the Voronoi tessellation for this dataset. Anomalies are depicted because each bus has its own polygon. The statistical distribution of the visualized versus actual data does shift the mean slightly but does not bias against the tails, which allows for the identification of areas with high variability. Finally, the Voronoi structure is static; once calculated we can efficiently render new bus values onto the polygons. The potential limitation, however, is for small-scale maps, where multiple polygons may form a single pixel or small pixels that are not visually discernible. We would then be prone to the same over-plotting limitations that apply to glyphs, and a certain degree of smoothing would become necessary. Nonetheless, the Voronoi tessellation is able to both faithfully represent bus values while providing a full contour-like covering for quick assessment of network trends, making it a strong alternative to contour maps and glyph maps. 

The H3 and S2 tessellations offer promising alternatives to contour maps for their improved representation of the statistical data distribution and regular spatial units, but they are prone to the same issues of aggregating outlying values and failing to relay the structure of the network. And in our judgment, the S2 mapping is aesthetically noisy. Further research could try selecting the highest deviation value for each cell in order to preserve outliers.

Finally, the network-weighted contour map builds on the regular contour map by allowing for discontinuities that convey the network structure, however is prone to smoothing some of the larger outliers based on the choice of n and k, which might be inevitable in small-scale maps. Further research should compare the networked contour map side by side with the traditional contour map for given values of $k$ and $n$ to control for effective size of the kernel.

As mentioned, these results hold true for this dataset of about 24,000 buses in an urban area with a relatively high but variable density. Lower density regions or smaller grids, where there are many pixels per bus, can afford to depict each bus with methods such as glyphs or Voronoi polygons, which preserves the full variability of the original data. However, larger grids with 100,000+ buses may have more buses than pixels, and are forced to aggregate the data in some way. Here, we can look to tiling or to contour maps. Using a small kernel or high resolution tiles helps preserve the variability of the original dataset. Using tiles or networked contour maps allows for discrete boundaries in the visualization, helping to preserving the spatial discontinuities in the data. Future research should evaluate these visualization methods on a large dataset that necessitates aggregation.  

Additionally, future research should also explore adding in time-series data to the network through animations or other methods. Further topological information could be conveyed by overlaying multiple types of visualizations or by designing linked views. Finally, follow up research should conduct a human factors study by surveying the preferences of power systems users or by testing human readability of each of these visualizations. We hope to inspire further investigation into the proposed methods and new alternative methods to better display power systems data in an intuitive, efficient, and beautiful way.

\section{Conclusion}
In summary, we took on the challenge of visualizing the current state of an urban electrical grid with over 24,000 buses. Instead of the standard method of colored contour maps, we tried alternative methods that vary the tile shape and size, add multiple layers, and consider the network topology. We evaluated the effectiveness of each of these visualizations by their ability to accurately depict the data with the same statistical distribution and to do so with computational efficiency. We found that a Voronoi tessellation works well for this dataset that has more pixels than buses, and that a networked contour map holds promise this and larger datasets. We look forward to further research that considers time series, linked views, and a human factors study. 

\acknowledgments{%
	The authors would like to thank Kristi Potter for her contribution to this research. This work was authored by the National Renewable Energy Laboratory, managed and operated by Alliance for Sustainable Energy, LLC for the U.S. Department of Energy (DOE) under Contract No. DE-AC36-08G028308. Funding provided by the U.S. Department of Energy Office of Energy Efficiency and Renewable Energy. The views expressed in the article do not necessarily represent the views of the DOE or the U.S. Government. The U.S. Government retains and the publisher, by accepting the article for publication, acknowledges that the U.S. Government retains a nonexclusive, paid-up, irrevocable, worldwide license to publish or reproduce the published form of this work, or allow others to do so, for U.S. Government purposes.
}

\bibliographystyle{abbrv-doi-hyperref}

\bibliography{references}

\end{document}